\documentclass[11pt, a4paper, Physsubmission, Phys]{SciPost}

\hypersetup{
    colorlinks,
    linkcolor={red!50!black},
    citecolor={blue!50!black},
    urlcolor={blue!80!black}
}

\usepackage[bitstream-charter]{mathdesign}
\DeclareSymbolFont{usualmathcal}{OMS}{cmsy}{m}{n}
\DeclareSymbolFontAlphabet{\mathcal}{usualmathcal}

\begin{document}

\begin{center}{\Large \textbf{
A measurement of the proton plus helium spectrum of cosmic
rays in the TeV region with HAWC\\
}}\end{center}

\begin{center}
J. C. Arteaga-Vel\'{a}zquez\textsuperscript{1,*}
on behalf of the HAWC collaboration.
\end{center}
\begin{center}
{\bf 1} Instituto de Física y Matemáticas, Universidad Michoacana de San Nicolás de Hidalgo
\\

* juan.arteaga@umich.mx
\\A complete list of HAWC's authors is available at: https://www.hawc-observatory.org/collaboration/
\end{center}



\definecolor{palegray}{gray}{0.95}
\begin{center}
\colorbox{palegray}{
  \begin{tabular}{rr}
  \begin{minipage}{0.1\textwidth}
    \includegraphics[width=30mm]{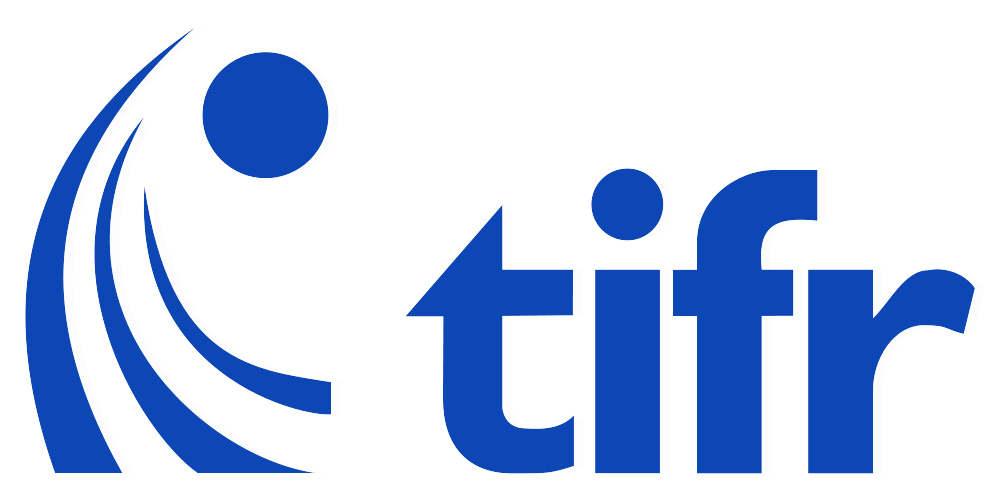}
  \end{minipage}
  &
  \begin{minipage}{0.85\textwidth}
    \begin{center}
    {\it 21st International Symposium on Very High Energy Cosmic Ray Interactions (ISVHE- CRI 2022)}\\
    {\it Online, 23-27 May 2022} \\
    \doi{10.21468/SciPostPhysProc.?}\\
    \end{center}
  \end{minipage}
\end{tabular}
}
\end{center}

\section*{Abstract}
{\bf
HAWC is an air-shower detector designed to study TeV gamma and cosmic rays. The
observatory is composed of a $22000 \, m^2$ array of $300$ water Cherenkov tanks ($4.5 \, m$ deep x $7.3 \, m$ diameter) with $4$ photomultipliers (PMT) each. The instrument registers the number of hit PMTs, the timing information and the total charge at the PMTs during the event. From these data, shower observables such as the arrival direction, the core position at ground, the lateral age and the primary energy are estimated. In this work, we study the distribution of the shower
age vs the primary energy of a sample of shower data collected by HAWC from June 2015 to June 2019 and employ a shower-age cut based on predictions of QGSJET-II-04 to separate a subsample of events dominated by H and He primaries. Using these data and a dedicated analysis, we reconstruct the cosmic ray spectrum of H+He from $6$ to $158$ TeV, which shows the presence of a softening at around $24$ TeV with a statistical significance of $4.1\sigma$. 
}


\section{Introduction}
\label{sec:intro}
 
  The $10 - 100$ TeV energy region of the cosmic-ray energy spectrum has been recently the target of both direct and indirect experiments due to the lack of in-depth studies on the composition and spectrum of cosmic-ray particles in this energy range. This was mainly the result of experimental  limitations of direct and indirect detection techniques. Recent progress on the matter has allowed to overcome such constraints allowing to boost detailed studies on the properties of such particles. As an example, on the side of indirect cosmic-ray detection, we have some recent studies from HAWC on the energy spectrum and composition of cosmic rays at tens of TeV \cite{Hawc2017, Hawc2022}. Composition measurements of cosmic rays from $10$ TeV to $100$ TeV with extensive air-shower (EAS) experiments are scarce.  Just a few EAS experiments have reported  data on cosmic-ray composition in this energy regime. In particular,  ARGO-YBJ performed measurements on the spectrum of the H$+$He mass group  \cite{Argo2012, Argo2015},  EAS-TOP with MACRO, on the intensity of H, He and CNO primaries \cite{Eastop2004}, KASCADE, on the flux of p primaries \cite{Kascade2004}, MAGIC, on the intensity of protons \cite{Magic2021}, and HESS \cite{Hess2007} and VERITAS \cite{Veritas2018}, on  the spectrum of Fe nuclei. Now, HAWC has joined  these efforts with a measurement of the cosmic-ray spectrum of H$+$He between $6$ and $158$ TeV  \cite{Hawc2022}. In this work, we will present a brief description of this analysis. The procedure is simple: we selected a subsample of HAWC events enriched with H and He nuclei based on a cut on the shower age that is derived from MC simulations using FLUKA/QGSJET-II-04. Then, we unfolded the energy distribution of the data subsample and, finally, we corrected it for the contamination of heavy elements ($Z > 2$) and for losses in the trigger and reconstruction efficiency. With MC studies, we found that the H$+$He mass group can be easily separated with this procedure due to its large relative abundance in the energy interval from $10$ to $100$ TeV. For this reason, we selected this group for the analysis. More details and discussions about this study and its results can be found in \cite{Hawc2022}. 
 
  \section{HAWC observatory, data and simulations}
  \label{sec:hawc}

   HAWC is an air-shower Cherenkov detector dedicated to measure gamma and cosmic rays in the energy interval from $100$ GeV up to $100$ TeV, and even up to $1$ PeV in case of cosmic-ray primaries.  It is located in the east-central part of Mexico, at an altitude of $4100 \, \mbox{m}$  at the Sierra Negra Volcano \cite{Hawc2017crab}. The HAWC central detector is composed of an array of $300$ water Cherenkov detectors ($7.3 \, \mbox{m}$ diameter $\times$ $4.5 \, \mbox{m}$ deep), which covers a surface of $22000 \, \mbox{m}^{2}$. Each of these detector units has $4$ PMTs, which provide data on the arrival times and the effective charge  of the front of EAS events. From these measurements a dedicated reconstruction program estimates different air shower observables as the core location, arrival direction, the lateral shower profile, among others \cite{Hawc2017crab, Hawc2019}. For this study, of particular interest are the lateral shower age $s$ and the estimated primary energy $E_{rec}$ of the event. The shower age is obtained event by event from a $\chi^2$ fit with a Nishimura-Kamata-Greisen (NKG) like function to the measured lateral charge distribution \cite{Hawc2019}. 
   On the other hand, the primary energy is estimated with a maximum likelihood procedure that compares the observed lateral charge profile with Monte Carlo (MC) templates for the lateral charge distributions   \cite{Hawc2017}. The templates were produced with CORSIKA \cite{corsika} and the hadronic interaction models FLUKA \cite{fluka} and QGSJET-II-04 \cite{qgsjet} using proton-induced showers and different zenith angle and energy bins, which cover the zenith angle interval $\theta < 60^\circ$ and the primary energy range  $E = 70 \, \mbox{GeV} - 1.4 \, \mbox{PeV}$, respectively. 
   
   For the present analysis, we employed HAWC data collected between  June 11, 2015 to June 3, 2019, corresponding to an effective time of $T_{eff} = 3.74 \, \mbox{years}$. To reduce the  systematic errors in our study, we applied the following data selection criteria: we employed showers with zenith angles smaller than $\sim 16.7^\circ$, which were successfully reconstructed,  that activated at least $40$ PMTs within a radius of $40 \,\mbox{m}$ from the shower core, which exhibited a fraction of hit PMTs greater or equal $0.2$ and reconstructed energy in the interval $\log_{10}(E_{rec}/\mbox{GeV}) = [3.5, 5.5]$. These cuts were also used on our MC simulations.
   
   The MC data sets for our analysis were produced with CORSIKA using FLUKA and QGSJET-II-04  for hadronic energies $<  80 \, \mbox{GeV}$ and $\geq 80 \, \mbox{GeV}$, respectively. We simulated eight primary nuclei with energies from $5 \, \mbox{GeV}$ to $3 \,\mbox{PeV}$ and an $E^{-2}$ spectrum, which were weighted to follow broken-power law spectra  \cite{Hawc2017, Hawc2022} that were fit to data from AMS-2 \cite{ams}, CREAM \cite{cream} and PAMELA \cite{pamela}. 
   
   After selection cuts, we kept $1.6 \times 10^{10}$ showers in the measured data set and
   $3.43 \times 10^5$ events in the MC simulations.
   Above $10 \, \mbox{TeV}$, using our MC data set, we estimated that the selected data sample has energy (in logarithmic scale) and core  resolutions smaller than $0.26$ and  $15 \, \mbox{m}$, respectively, and a pointing resolution better than $0.55^\circ$.

 \section{Analysis procedure}
  \label{sec:analysis}
  
\begin{figure}[!b]
	\begin{center}
		\hspace*{0cm}\begin{tabular}{ c c }
		
		\includegraphics[width=0.47\linewidth,height=0.36 \linewidth]{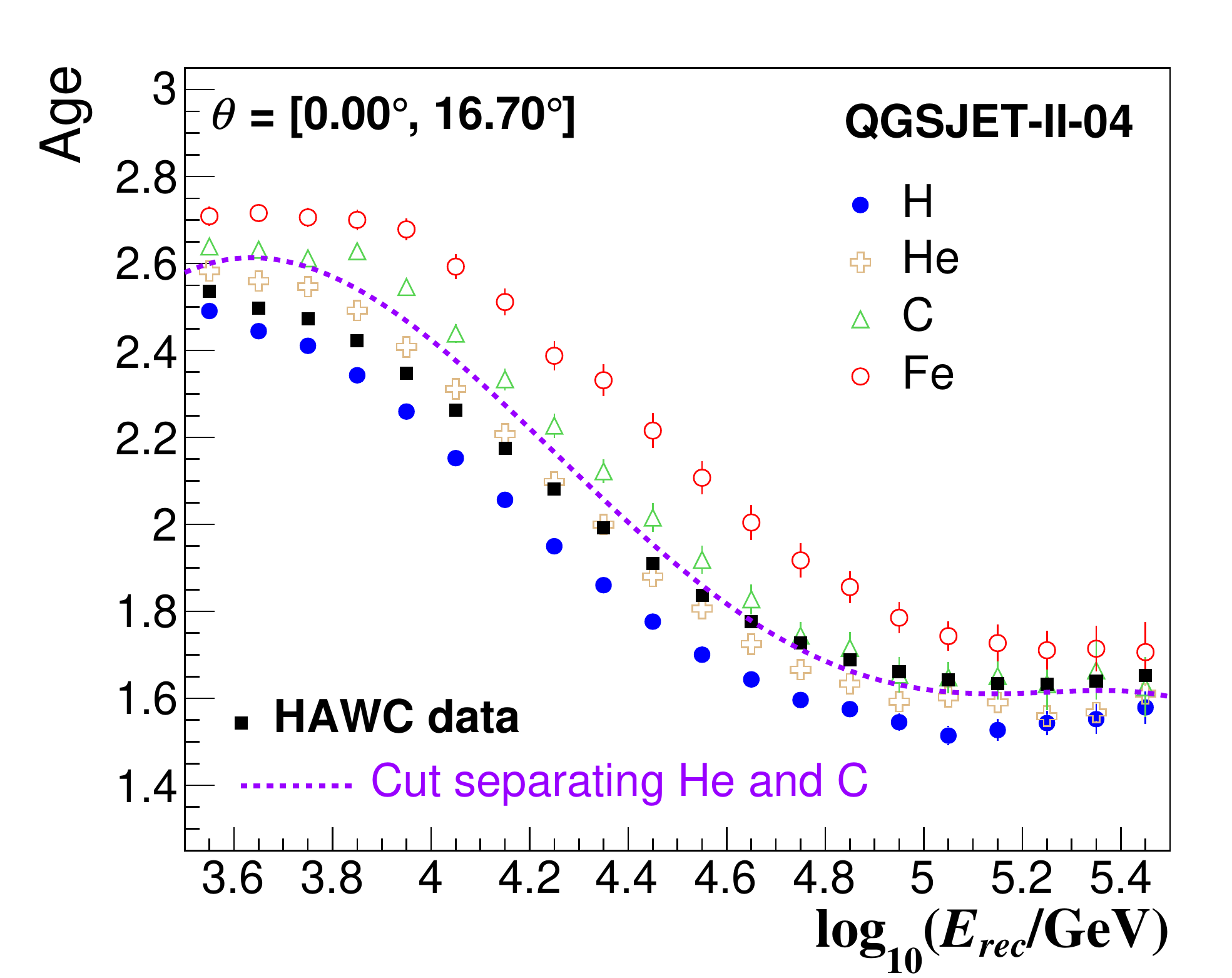}  &
		\includegraphics[width=0.50\linewidth,height=0.36 \linewidth]{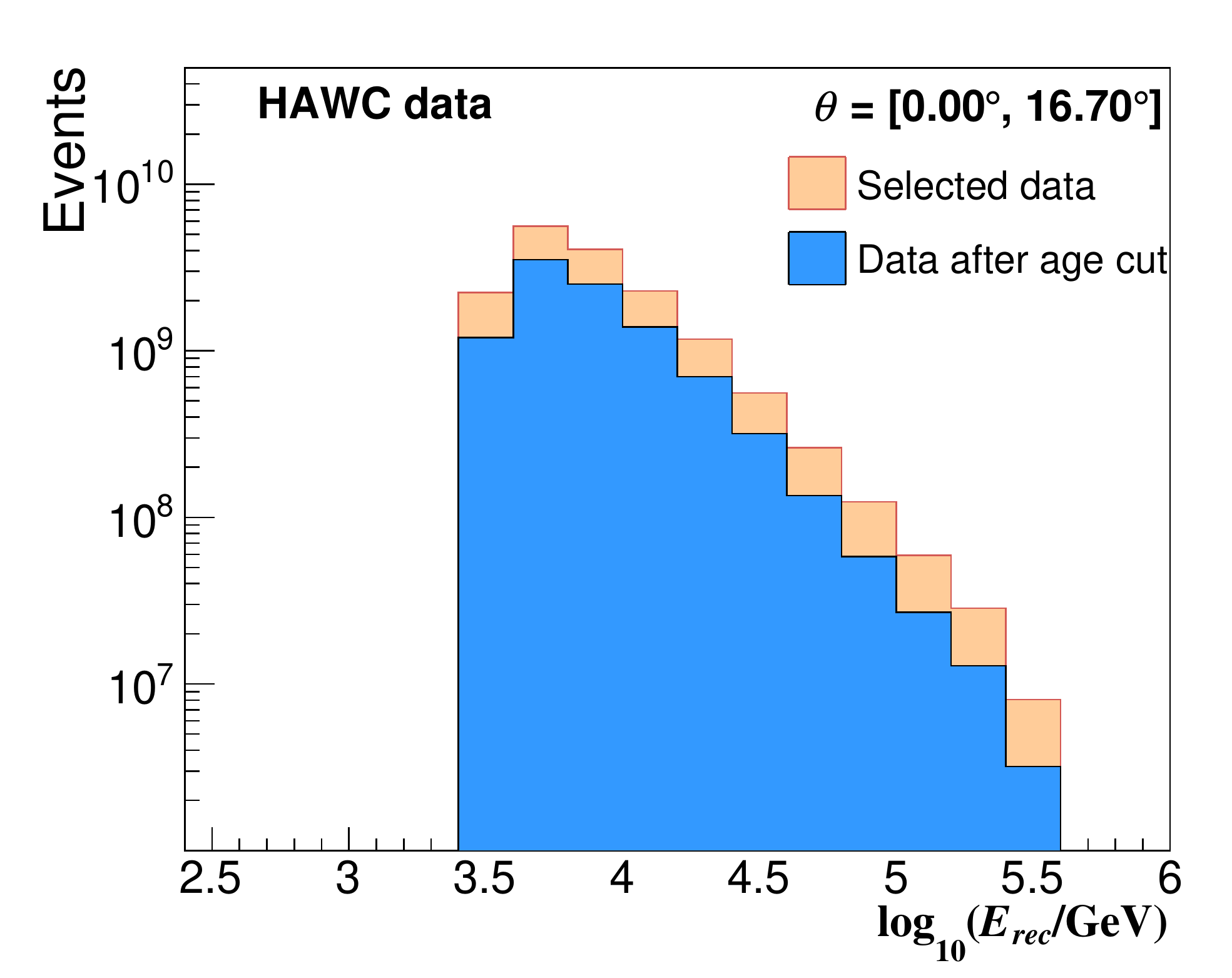}
	\end{tabular}
	\end{center}
	\caption{\textit{\small Left: FLUKA/QGSJET-II-04 preditions for the mean shower age vs the reconstructed primary energy of different cosmic-ray nuclei: H (filled circles), He (crosses), C (triangles) and Fe (open circles). The segmented line represents the shower-age cut used for the selection of our subsample of young showers for the analysis. The mean shower age for HAWC data is shown with squares \cite{Hawc2022}. Right: The raw energy histogram for the data after applying quality cuts (orange) is compared with the corresponding histogram for the subsample of young EAS obtained after using the shower-age cut \cite{Hawc2022}. }} 
	\label{Fig_1_2}
  \end{figure}

  To start the analysis, we applied an energy-dependent cut on the lateral shower age  to select a subsample of events dominated by protons and helium primaries. Using our MC simulations, we put the cut between the mean predictions for He and C nuclei and we only kept young EAS with $s$ values smaller than this cut, as such events are most likely produced by H and He primaries. The MC predictions for the mean shower age of EAS induced by different primary nuclei are shown in the left plot of Fig.~\ref{Fig_1_2} as a function of the reconstructed energy. The shower-age cut applied on the analysis is also displayed. From MC simulations and our nominal composition model, the purity of the subsample is expected to be greater than $82\%$ in each bin of $E_{rec}$ using our selection criterion. The energy histogram of the experimental data sample after applying the shower cut is shown on  the right plot of Fig.~\ref{Fig_1_2}. It is compared with the energy histogram before applying the age cut on the selected data. 
 
\begin{figure}[!t]
	\begin{center}
		\hspace*{0cm}\begin{tabular}{ c c }
		\includegraphics[width=0.50\linewidth,height=0.36 \linewidth]{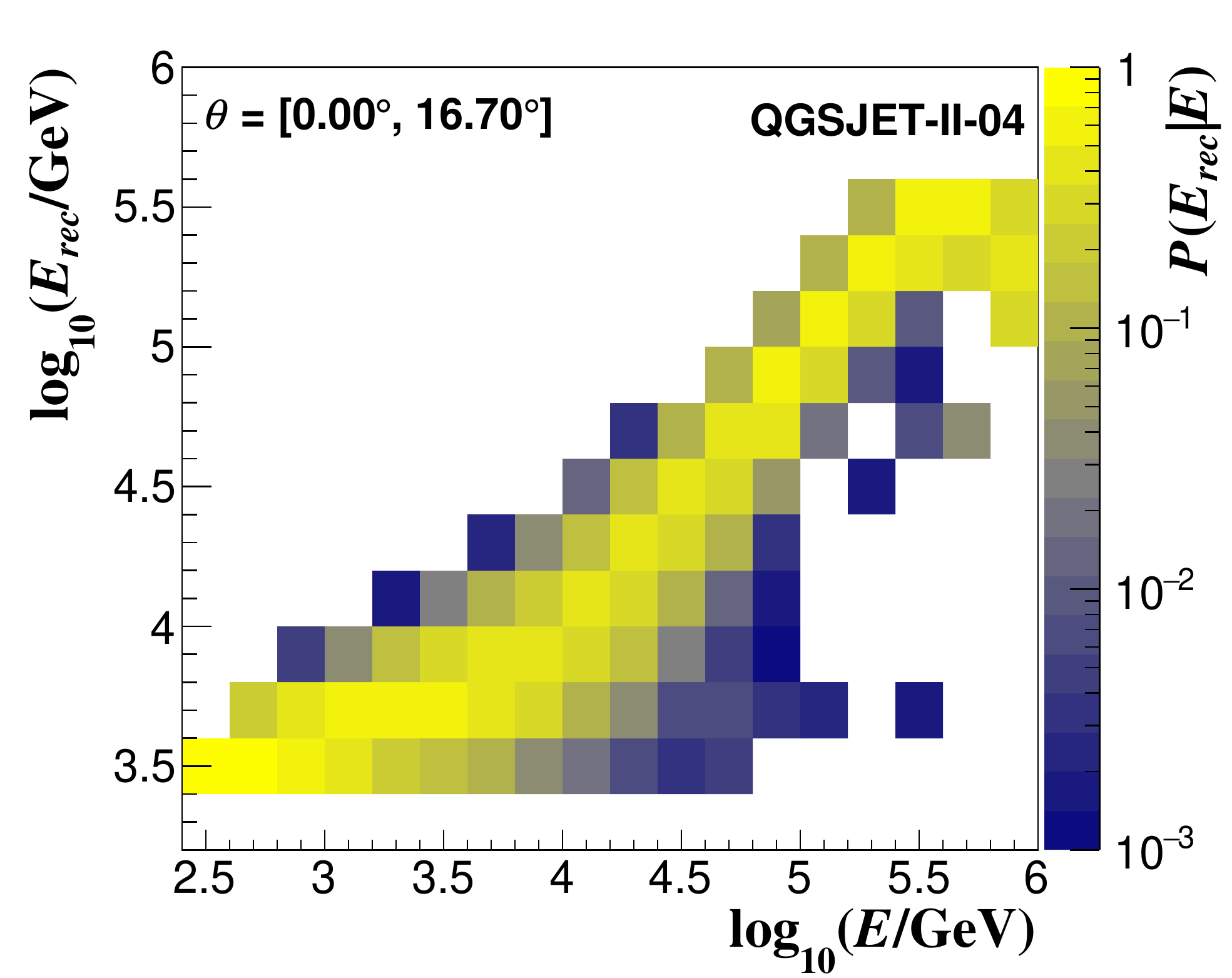}  &
		\includegraphics[width=0.47\linewidth,height=0.36 \linewidth]{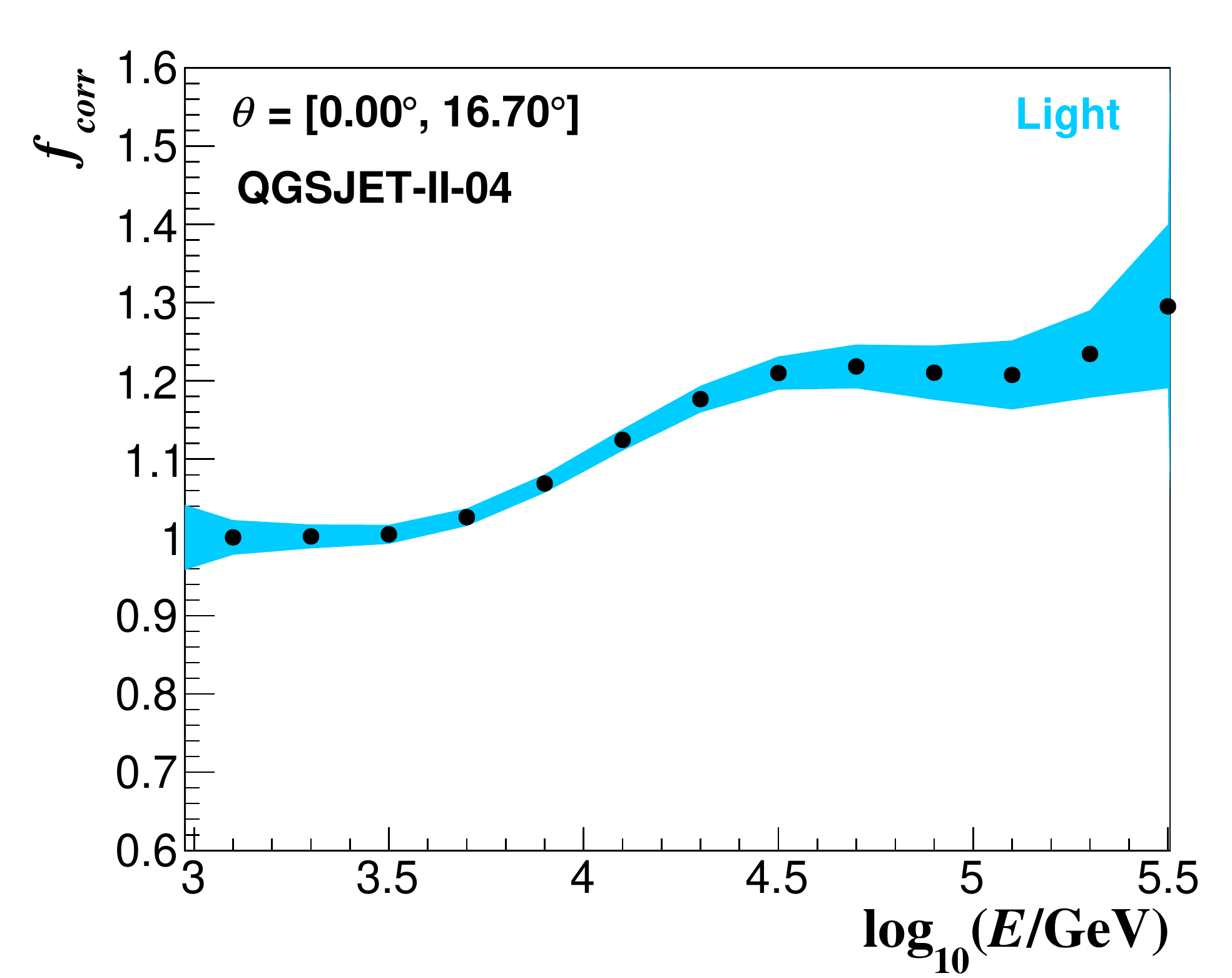}
	\end{tabular}
	\end{center}
	\caption{\textit{\small Left: The response matrix for the subsample of young EAS calculated with QGSJET-II-04 and our cosmic-ray composition model \cite{Hawc2022}.  Right: $f_{corr}$ vs the true primary energy $E$ according to MC simulations. The error band represents statistical uncertainties \cite{Hawc2022}.}} 
	\label{Fig_3_4}
  \end{figure}

  The subsample of young EAS has almost $9.9 \times 10^{9}$ events and its histogram is not corrected for migration effects yet. For this reason, we applied the method of Bayesian unfolding \cite{bayes_unfold}. The iteration depth was set at the minimum of the Weighted Mean Squared Error \cite{kascade2}. The prior distribution was calculated from the spectrum for the H$+$He intensity in our nominal composition model. In addition, in every intermediate step during the unfolding procedure, but excluding the last one, we smoothed the energy spectrum using a fit with a broken power-law function \cite{bpl}.  The response matrix employed for unfolding the raw energy distribution was obtained using MC simulations and our nominal composition model (see left plot of Fig.~\ref{Fig_3_4}).

  In a next step, we calculated a factor $f_{corr}$ to correct the unfolded result $N(E)$ for the contamination of heavy elements (c.f. right plot of Fig.~\ref{Fig_3_4}). This quantity is computed from MC simulations for different true energy ($E$) bins as the inverse of the proportion of H and He nuclei in the subsample of young showers. Then, to correct for losses of trigger and reconstruction efficiency, we calculated the effective area for the H$+$He mass group $A_{eff}^{H+He}$ with MC simulations using our nominal composition model as a function of  $E$ (see left plot of Fig.~\ref{Fig_5_6}). Finally, the energy spectrum of H$+$He nuclei was estimated as 
  \begin{equation}
    \Phi(E) = \frac{N(E)}{\Delta E \, T_{eff} \,  \Delta \Omega \, \, f_{corr}(E) \, \, A_{eff}^{H+He}(E)}, 
  \end{equation}
  where $\Delta \Omega = 0.27 \, \mbox{sr}$
  is the solid-angle interval of observation.

  \section{Results}
  
\begin{figure}[!t]
	\begin{center}
		\hspace*{0cm}\begin{tabular}{ c c }
		
		\includegraphics[width=0.42\linewidth,height=0.36 \linewidth]{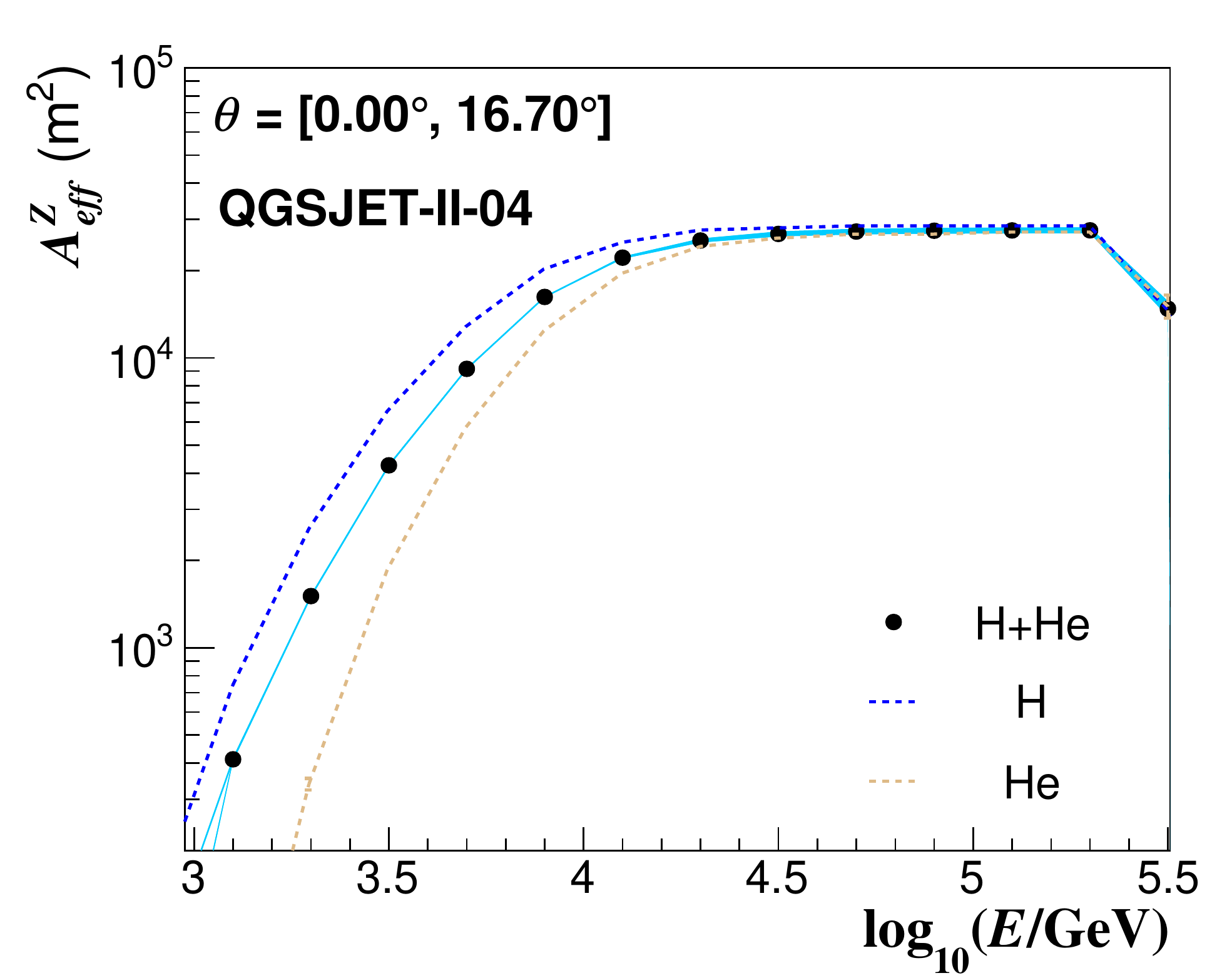}  &
		\includegraphics[width=0.55\linewidth,height=0.36 \linewidth]{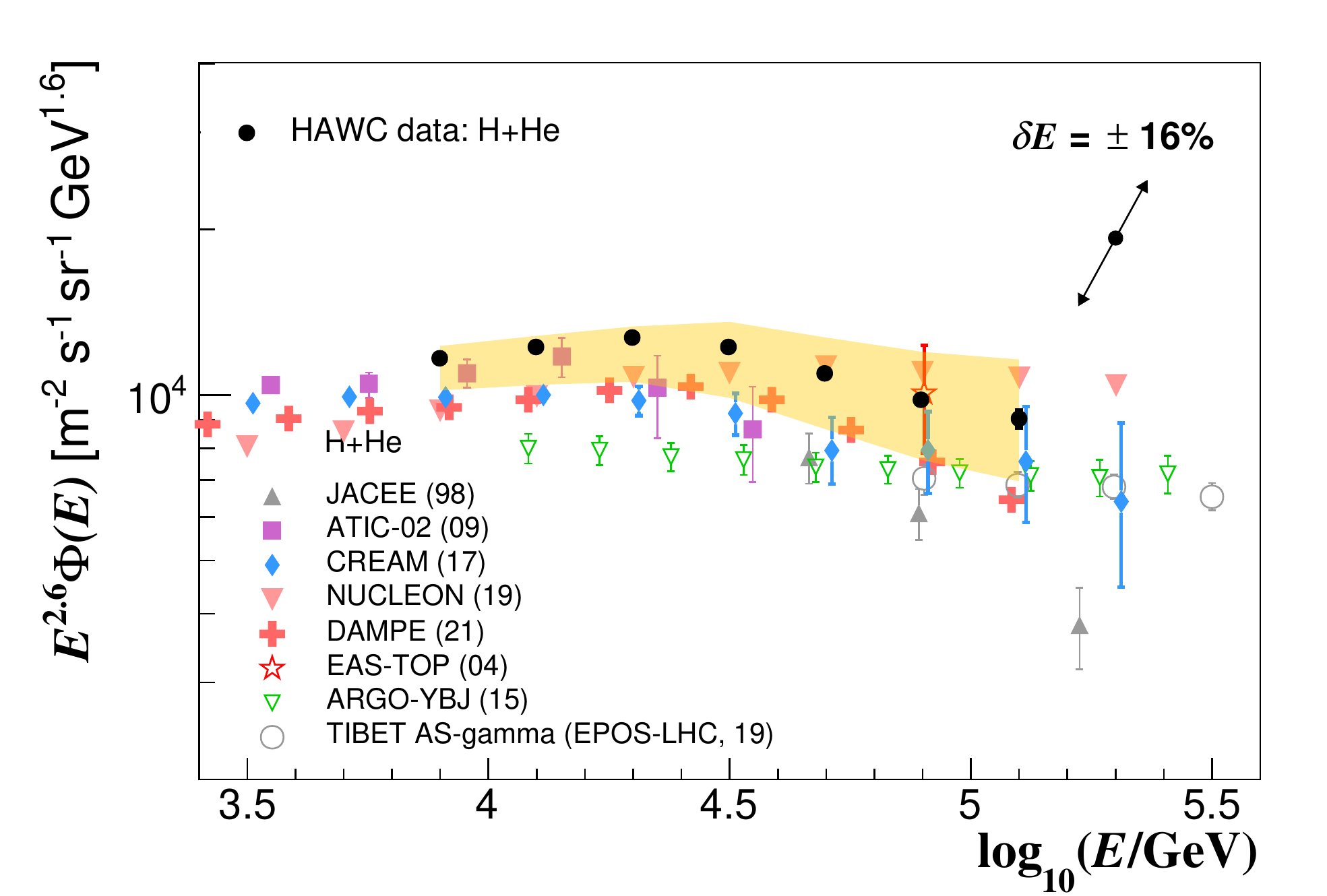}
	\end{tabular}
	\end{center}
	\caption{\textit{\small Left: The effective area for H$+$He primaries (data points)  as a function of the true primary energy as estimated with MC simulations \cite{Hawc2022}. It is compared with the effective area expected for pure protons (blue dashed line) and He nuclei  (orange dashed line). Statistical uncertainties are shown with an error band. Right: The energy spectrum of  H$+$He cosmic rays as measured with HAWC (black data points) in comparison with TeV measurements of the same mass group from direct and indirect cosmic-ray experiments  \cite{Hawc2022} (see text for a description). The error band for HAWC data represents the systematic uncertainties, while the vertical error bars, statistical errors. For the other data points, only statistical errors are displayed.}} 
	\label{Fig_5_6}
  \end{figure}

  The energy spectrum of H$+$He nuclei is shown in the right plot of Fig.~\ref{Fig_5_6} along with its statistical and systematic errors. Inside the interval $\log_{10}(E/\mbox{GeV}) = [3.8, 5.2]$, statistical errors are smaller than $3.8\%$. They include the statistical uncertainties due to the limited statistics of the MC simulations and the statistics of the measured data, both added in quadrature. On the other hand, systematic errors are found within $-22.6\%$ and $+28.9\%$. They incorporate uncertainties from the modelling of the PMTs, the high-energy hadronic interaction model (for which we also used EPOS-LHC \cite{eposlhc} simulations), the relative abundances of cosmic rays (we employed the Polygonato model \cite{polygonato}, three additional composition models calibrated with direct data from ATIC-2 \cite{atic07}, JACEE \cite{jacee} and MUBEE \cite{mubee}, respectively, and an additional model for which we varied the intensity of the heavy component  in our nominal composition model to match the  observed abundance of heavy primaries from an analysis of the efficiency of the shower-age cut \cite{Hawc2022}), the seed spectrum in the unfolding procedure (changing the prior spectrum by uniform and $E^{-1.5}$ distributions), the method of unfolding (using also the Gold's algorithm \cite{Gold_unfold}), the smoothing procedure (employing a fit with a 5th degree polynomial and the 353HQ-twice algorithm \cite{353HQ}  of ROOT \cite{root}, respectively), the corrected effective area and the position of the shower-age cut (putting the cut at the curves for the mean age of He and C primaries, respectively). The largest contributions to the systematic error are due to uncertainties in the PMT modelling (it is found within  $-10.6\%/+28.5\%$), the high-energy hadronic interaction model (it contributes with an error between $-10.9\%$ and $-3.7\%$) and the cosmic-ray composition (with errors between $-17.3\%$ and $+2.1\%$). The sum in quadrature of the remaining uncertainty sources ranges from $-7\%$ to $+5.2\%$. In all cases the shape of the spectrum is preserved.

  In the spectrum of HAWC for H$+$He nuclei, we observe the presence of a break at some tens of TeV. A fit of the spectrum with a broken power-law expression \cite{bpl}
  \begin{equation}
      \Phi(E) = \Phi_{0} E^{\gamma_1} \left[ 1  + \left( \frac{E}{E_{0}} \right)^\varepsilon \right]^{(\gamma_2 - \gamma_1)/\varepsilon},
  \end{equation}
  shows the existence of a softening at around $E_{0} = 24.0^{+3.6}_{-3.1} \, \mbox{TeV}$ associated with a change in the spectral index from $\gamma_1 = -2.51 \pm 0.02$ to $\gamma_2 = -2.83 \pm 0.02$. The fit gave a chi-squared $\chi^2 = 0.26$ for $2$ degrees of freedom. A statistical analysis based on the difference between the $\chi^2$ of this fit with that obtained with a power-law formula as a test statistics yielded a statistical significance of $4.1\sigma$ for the feature.

  To end, the HAWC result is compared with measurements of direct and other indirect cosmic-ray experiments in the TeV energy range in the right plot of  Fig.~\ref{Fig_5_6}. We observe that HAWC results confirm previous hints from ATIC-2 \cite{atic09}, CREAM I-III \cite{cream17} and NUCLEON \cite{nucleon19} about the existence of a break in the spectrum of H$+$He cosmic rays at tens of TeV. This result is also strengthen by recent DAMPE data \cite{dampe21}. We also see that our result is in agreement with ATIC-2 and NUCLEON within systematic uncertainties, but it seems to be above  DAMPE (at low energies) and CREAM data. The HAWC spectrum is also above the ARGO-YBJ \cite{Argo2015} and TIBET AS$-\gamma$ \cite{Tibet19} measurements, but in agreement with EAS-TOP data \cite{Eastop2004}. 
  
  \section{Conclusion}
   
   A dedicated analysis of cosmic ray composition with HAWC has allowed to measure the spectrum of H$+$He of cosmic rays in the energy range $E = [6, 158] \, \mbox{TeV}$. The spectrum shows a softening at $ 24.0^{+3.6}_{-3.1} \, \mbox{TeV}$ with a statistical significance of $4.1\,\sigma$. This analysis shows the potential of high-altitude water Cherenkov observatories like HAWC for composition studies of cosmic rays.

   \footnotesize

  \section*{Acknowledgements}
 The main list of acknowledgements can be found under the following link:
 \href{https://www.hawc-observatory.org/collaboration/}{https://www.hawc-observatory.org/collaboration}. In addition, J.C.A.V. also wants to thank the partial support from CONACYT grant A1-S-46288 and the Coordinaci\'{o}n de la Investigación Científica de la Universidad Michoacana.


\begin{thebibliography}{10}
\providecommand{\url}[1]{\texttt{#1}}
\providecommand{\urlprefix}{URL }
\expandafter\ifx\csname urlstyle\endcsname\relax
  \providecommand{\doi}[1]{doi:\discretionary{}{}{}#1}\else
  \providecommand{\doi}{doi:\discretionary{}{}{}\begingroup
  \urlstyle{rm}\Url}\fi
\providecommand{\eprint}[2][]{\url{#2}}

\bibitem{Hawc2017}
R.~Alfaro, HAWC~collaboration,
\newblock Phys. Rev. D \textbf{96}, 122001 (2017).

\bibitem{Hawc2022}
A.~Albert, HAWC~Collaboration,
\newblock Phys. Rev. D \textbf{105}, 063021 (2015).

\bibitem{Argo2012}
B.~Bartoli, ARGO-YBJ~Collaboration,
\newblock Phys. Rev. D \textbf{85}, 092005 (2012).

\bibitem{Argo2015}
B.~Bartoli, ARGO-YBJ~Collaboration,
\newblock Phys. Rev. D \textbf{91}, 112017 (2015).

\bibitem{Eastop2004}
M.Aglietta, EAS-TOP~Collaboration and the MACRO~Collaboration,
\newblock Astropart. Phys. \textbf{21}, 223 (2004).

\bibitem{Kascade2004}
T.~Antoni, KASCADE~Collaboration,
\newblock The Astrophys. J. \textbf{612}, 914 (2004).

\bibitem{Magic2021}
P.~Temnikov, MAGIC~Collaboration,
\newblock Proceedings of Science Vol. 375 \textbf{231- 37th ICRC (Berlin,
  Germany)} (2021).

\bibitem{Hess2007}
F.~Aharonian, HESS~Collaboration,
\newblock Phys. Rev. D \textbf{75}, 042004 (2007).

\bibitem{Veritas2018}
A.~Archer, VERITAS~Collaboration,
\newblock Phys. Rev. D \textbf{98}, 022009 (2018).

\bibitem{Hawc2017crab}
A.~U. Abeysekara, HAWC~collaboration,
\newblock The Astrophys. J. \textbf{843}, 39 (2017).

\bibitem{Hawc2019}
A.~U. Abeysekara, HAWC~collaboration,
\newblock The Astrophys. J. \textbf{881}(2), 134 (2019).

\bibitem{corsika}
D.~Heck, J.~Knapp, J.~N. Capdevielle, G.~Schatz and T.~Thouw,
\newblock \emph{{CORSIKA: A Monte Carlo Code to Simulate Extensive Air
  Showers}},
\newblock Report FZKA \textbf{6019}(11) (1998).

\bibitem{fluka}
G.~Battistoni \emph{et~al.},
\newblock \emph{{The FLUKA code: Description and benchmarking}},
\newblock In \emph{AIP Conference proceedings}, vol. 896, pp. 31--49. American
  Institute of Physics (2007).

\bibitem{qgsjet}
S.~Ostapchenko,
\newblock Phys. Rev. D \textbf{83}(1), 014018 (2011).

\bibitem{ams}
M.~Aguilar, AMS~collaboration,
\newblock Phys. Rev. Lett. \textbf{115}(21), 211101 (2015).

\bibitem{cream}
Y.~Yoon, CREAM~collaboration,
\newblock The Astrophysical Journal \textbf{728}(2), 122 (2011).

\bibitem{pamela}
O.~Adriani, PAMELA~collaboration,
\newblock Science \textbf{332}(6025), 69 (2011).

\bibitem{bayes_unfold}
G.~D'Agostini,
\newblock NIMA \textbf{362}(2), 487 (1995).

\bibitem{kascade2}
T.~Antoni, KASCADE~collaboration,
\newblock Astropart. phys. \textbf{24}(1-2), 1 (2005).

\bibitem{bpl}
S.~V. Ter-Antonyan and L.~S. Haroyan,
\newblock \emph{About $\mbox{EAS}$ size spectra and primary energy spectra in the knee
  region},
\newblock \eprint{https://doi.org/10.48550/arXiv.hep-ex/0003006}.

\bibitem{eposlhc}
T.~Pierog, I.~Karpenko, J.~M. Katzy, E.~Yatsenko and K.~Werner,
\newblock Phys. Rev. C \textbf{92}, 034906 (2015).

\bibitem{polygonato}
J.~Hoerandel,
\newblock Astropart. Phys. \textbf{19}(2), 193 (2003).

\bibitem{atic07}
A.~Panov, ATIC-2 Collaboration,
\newblock Bull. Russ. Acad. Sci. Phys. \textbf{71}, 494 (2007).

\bibitem{jacee}
Y.~Takahashi, JACEE Collaboration,
\newblock Nucl. Phys. B, Proc. Suppl. \textbf{60}, 83 (1998).

\bibitem{mubee}
V.~I.~Zatsepin, MUBEE Collaboration,
\newblock In D.~A. Leahy, R.~B. Hickws and D.~Venkatesan, eds.,
  \emph{Proceedings of the 23rd ICRC (Calgary, Canada), Vol. 2}, No. 13. World
  Scientific (1993).

\bibitem{Gold_unfold}
R.~Gold,
\newblock \emph{An iterative unfolding method for response matrices},
\newblock Report No ANL-6984  (1964).

\bibitem{353HQ}
J.~Friedman,
\newblock \emph{Data analysis techniques for high energy particle physics},
\newblock In \emph{Proceedings of the 3rd CERN School of Computing (Norway),
  CERN, Geneva}, p. 271 (1974).

\bibitem{root}
R.~Brun and F.~Rademakers,
\newblock Nucl. Instrum. Methods A \textbf{389}, 81 (1997).

\bibitem{atic09}
A.~Panov, ATIC-2 Collaboration,
\newblock Bull. Russ. Acad. Sci. Phys. \textbf{73}, 564 (2009).

\bibitem{cream17}
Y.~Yoon, CREAM Collaboration,
\newblock The Astrophys. J. \textbf{839}, 5 (2017).

\bibitem{nucleon19}
E.~V. Atkin, NUCLEON Collaboration,
\newblock Astronomy Reports \textbf{63}, 66 (2019).

\bibitem{dampe21}
F.~Alemanno, DAMPE Collaboration,
\newblock Proceedings of Science Vol. 395 \textbf{117 - 37th ICRC (Berlin,
  Germany)} (2021).

\bibitem{Tibet19}
M.~Amenomori, Tibet AS$-\gamma$ Collaboration,
\newblock EPJ Web Conf. \textbf{208}, 03001 (2019).

\end{thebibliography}
\end{document}